\documentclass[reprint,amsmath,amssymb, prb, aps, showpacs]{revtex4-1}

\usepackage{amssymb}
\usepackage{amsmath}
\usepackage{hyperref}
\usepackage{color}
\usepackage{amsthm}
\usepackage{verbatim} 
\usepackage{graphicx}
\usepackage{ textcomp }

\newcommand{\FOP}{\mathcal{X}}
\newcommand{\SOP}{\mathcal{Y}}

\newcommand{\diag}{\text{diag}}
\newcommand{\Cl}{\text{Cl}}
\newcommand{\smat}{\left(\begin{smallmatrix}}
\newcommand{\stam}{\end{smallmatrix}\right)}
\newcommand{\Integer}{\mathbb{Z}}
\newcommand{\cZ}{\mathcal{Z}}

\begin{document}

\title{From symmetry-protected topological order to Landau order}
\author{Kasper Duivenvoorden}
 \email{Kasper@thp.uni-koeln.de}
\author{Thomas Quella}%
 \email{Thomas.Quella@uni-koeln.de}
 \affiliation{Institute of Theoretical Physics, University of Cologne\\
 Z\"ulpicher Stra\ss{}e 77, D-50937 Cologne, Germany 
}%

\date{\today}

\begin{abstract}
  Focusing on the particular case of the discrete symmetry group
  $\Integer_N\times\Integer_N$, we establish a mapping between
  symmetry protected topological phases and symmetry broken phases for
  one-dimensional spin systems. It is realized in terms of a non-local
  unitary transformation which preserves the locality of the
  Hamiltonian. We derive the image of the mapping for various phases
  involved, including those with a mixture of symmetry breaking and
  topological protection. Our analysis also applies to topological
  phases in spin systems with arbitrary continuous symmetries of
  unitary, orthogonal and symplectic type. This is achieved by
  identifying suitable subgroups $\Integer_N\times\Integer_N$ in all
  these groups, together with a bijection between the individual
  classes of projective representations.
\end{abstract}

\maketitle


\section{Introduction}

  Symmetry protected topological phases have received a lot of interest
  recently due to their characteristic properties such as the
  existence of massless boundary modes. Prominent examples are various
  types of topological insulators
  \cite{Kane:PhysRevLett.95.146802,Fu:PhysRevLett.98.106803} and spin
  systems such as the AKLT model.
  \cite{Affleck:PhysRevLett.59.799,Affleck:1987cy} In contrast to
  purely topological phases such as the fractional or integer quantum
  Hall effect, in these systems the robustness of boundary modes is
  directly tied to the presence of symmetries.

  For one-dimensional spin systems a complete classification of gapped
  symmetry protected topological and symmetry broken phases has been
  established in
  Refs.~\onlinecite{Chen:PhysRevB.83.035107,Schuch:1010.3732v3,Chen:PhysRevB.84.235128,Duivenvoorden:2012arXiv1206.2462D}.
  Restricting one's attention to on-site symmetries $G$ only, the
  phases are fully characterized by the spontaneous symmetry breaking
  of $G$ to a subgroup $K$, together with an element from the
  cohomology group $H^2(K,U(1))$. The latter labels the distinct
  classes of projective representations of $K$ and can be thought of
  as being a discrete topological invariant attached to edge
  modes of the system.\footnote{In the case of periodic boundary
    conditions, these edge modes arise virtually in the groundstate
    entanglement spectrum after the system is cut into two pieces.}

  In the present paper we shall focus on the particular symmetry group
  $\Integer_N\times\Integer_N$. It is the smallest Abelian group which
  exhibits up to $N$ distinct topological phases. At the same time,
  it allows us to study topological phases in combination with the
  phenomenon of spontaneous symmetry breaking
  if $N$ has non-trivial divisors. While information about the latter
  can be inferred from suitable local Landau order parameters, the
  detection of topological phases in 1D systems requires the use of
  non-local string order parameters (see, e.g., Refs.~\onlinecite{DenNijs:PhysRevB.40.4709,PhysRevLett.100.167202,Tu:2008JPhA...41O5201T,Haegeman:2012PhRvL.109e0402H,PhysRevB.86.125441,Duivenvoorden:2012arXiv1208.0697D}).
  As we shall see later, discrete groups of type
  $\Integer_N\times\Integer_N$ also play a distinguished role when
  extending our considerations to continuous symmetry groups.

  Before we proceed, let us briefly review the specific case of the
  dihedral group $D_2=\Integer_2\times\Integer_2$ which historically
  arose in connection with the $SO(3)$ AKLT model.
  \cite{Affleck:PhysRevLett.59.799,Affleck:1987cy} It is well-known
  that the AKLT model realizes the Haldane phase of $S=1$ spin
  models. It exhibits topological order which can be detected using
  the non-local string order parameter suggested by Rommelse and
  Den\,Nijs. \cite{DenNijs:PhysRevB.40.4709} Soon after, it was
  discovered that the presence of topological order could be
  interpreted as the spontaneous breaking of a ``hidden''
  $\Integer_2\times\Integer_2$-symmetry which is related to the
  occurrence of two spin 1/2 degrees of freedom at the edges of the
  chain. \cite{raey} This symmetry breaking becomes manifest after a
  non-local unitary transformation which maps the original string
  order parameter to a local Landau order parameter. This mapping can also be understood as disentangler,
  \cite{PhysRevB.83.104411} transforming the entangled AKLT state into a product state. It is known that
  this observation extends to AKLT chains based on higher integer
  spins $S$. \cite{0953-8984-4-36-019} However, it only became clear
  later that the topological protection does in fact not rely
  on the full $SO(3)$ symmetry but that it can already be achieved by
  restricting one's attention to a discrete subgroup
  $\Integer_2\times\Integer_2$.
  \cite{Pollmann:PhysRevB.81.064439,Pollmann:2012PhRvB..85g5125P} The
  non-trivial elements of this group can be thought of as rotations by
  $\pi$ around three mutually orthogonal axes. This group also appears in connection to the cluster state, where in a similar way it protects the topological order. \cite{Wonmin}

  In this paper, we generalize the previous ideas to arbitrary groups
  $\Integer_N\times\Integer_N$. In the process we face two
  difficulties: First of all, these groups allow for more complicated
  patterns of spontaneous symmetry breaking, giving rise to a whole
  hierarchy of phases. And second, a more refined version of string
  order parameter has to be used. Indeed, while for
  $\Integer_2\times\Integer_2$ one only has to distinguish between two
  topological phases (trivial and non-trivial), we now have to
  discriminate $N$ distinct phases which are labeled by a parameter
  $t\in\Integer_N$. It should be obvious that such a number cannot be
  extracted from a single expectation value since there is no reason
  why the latter should be quantized.

  As it turns out, the method of choice is to employ the selection
  rule procedure of Ref.~\onlinecite{PhysRevB.86.125441}.\footnote{Other
    discriminating string order parameters with a different scope of
    applicability have been suggested in Refs.~\onlinecite{Haegeman:2012PhRvL.109e0402H,Duivenvoorden:2012arXiv1208.0697D}.}
  Our analysis starts with a specific (string) order parameter
  $S(a,b)$ which depends on two parameters $a,b\in\Integer_N$ and
  which vanishes except if the selection rule $a+tb=0$ modulo $N$ is
  satisfied. While $S(a,b)$ is non-local for $b\neq0$ it becomes local
  for $b=0$. Determining $S(a,b)$ for various choices of $a$ and $b$,
  one can thus extract information about the topological phase $t$ and
  about the potential existence of spontaneous symmetry breaking. We
  then construct a non-local unitary transformation $U_N$ which maps
  $S(a,b)$ to $S(a,a+b)$. We analyze the implications of this mapping
  and find that purely topologically phases are mapped to
  symmetry breaking ones. The transformation $U_N$ thus allows us to
  reinterpret topological order in terms of standard local Landau
  order. More generally, we work out the effect of acting with
  $U_N$ on almost all phases of $\Integer_N\times\Integer_N$ spin
  chains, including those with a mixture of topological protection and
  spontaneous symmetry breaking to a subgroup
  $\Integer_r\times\Integer_r$.\footnote{These subgroups are those
    which are relevant for understanding the topological phases of
    systems with continuous symmetry.}

  As the original example of the $SO(3)$ AKLT model suggests, the
  results we derive may equally well be applied to the detection of
  topological order in systems with continuous symmetry
  groups.
  This is due to the fact that each continuous symmetry group
  which permits non-trivial topological phases (see
  Ref.~\onlinecite{Duivenvoorden:2012arXiv1206.2462D}), contains a
  non-trivial subgroup of the form $\Integer_N\times\Integer_N$ for a
  suitable choice of $N\geq2$. Moreover, the projective
  representations characterizing the topological phases with regard to
  either the continuous group or its discrete subgroup are in
  bijection.\footnote{This last statement is not valid for the series
    $PSO(4N)$. Here, one needs to work with a discrete Abelian
    subgroup involving more than two factors.}
  For all groups of unitary, orthogonal and symplectic type the
  relevant subgroups are constructed explicitly in
  Sect.~\ref{sec:continuoussymmetrygroups}.

  The paper is structured as follows. In Sect.~\ref{sec:preliminaries}
  we introduce the group $\Integer_N\times\Integer_N$ and we discuss
  its projective
  representations. Furthermore, we introduce the (string) order
  parameter $S(a,b)$ and explain how it can be used to characterize
  the distinct topological and symmetry broken phases of
  $\Integer_N\times\Integer_N$-invariant spin systems. The construction
  of the non-local unitary transformation $U_N$ which maps purely
  topological phases to symmetry broken ones is the content of
  Sect.~\ref{sec:NLTU}. In Sect.~\ref{sec:map} we analyze the fate of each
  individual phase under the action of $U_N$. Finally,
  Sect.~\ref{sec:continuoussymmetrygroups}
  discusses the implications of our results for continuous groups. We
  conclude with a brief summary and an outlook to future directions.

\section{\label{sec:preliminaries}Preliminaries}

  Different phases of one-dimensional quantum systems can either arise
  due to symmetry breaking, due to topology, or a combination
  thereof. We will consider systems with an on-site symmetry $G$, but
  we will not impose space-time symmetries such as translation
  invariance, time reversal or inversion symmetry. If the groundstate
  of the system breaks the symmetry to a subgroup $K\subset G$, then
  the possible topological phases are given by different classes of
  projective representations of $K$. \cite{
    Schuch:1010.3732v3,Chen:PhysRevB.84.235128}

  The group $\Integer_N\times\Integer_N$ has $N$ different
  projective classes. Let $R$ and $\tilde{R}$ be the generators of
  this group and let $R'$ and $\tilde{R}'$ be projective
  representations of these generators. The phases $R'^N = e^{i\theta}$
  and $\tilde{R}'^N = e^{i\tilde{\theta}}$ can be removed by a
  redefinition (gauge transformation) of $R'$ and
  $\tilde{R}'$. However, the phase $R'\tilde{R}'R'^{-1}\tilde{R}'^{-1}
  = e^{i\phi}$ is gauge invariant and determines the projective class
  of the corresponding representation. Moreover, $\phi$ is an integer
  multiple of $\frac{2\pi}{N}$ due to the cyclic property of the
  group $\Integer_N$. Thus the projective class $t\in\Integer_N$ of
  a representation of $\Integer_N\times\Integer_N$ can be obtained
  from the relation
\begin{align}\label{eq:define_t}
  R'\tilde{R}' = \omega^t \tilde{R}'R'\ \ .
\end{align} 
  Here, we used the abbreviation $\omega = \exp(2\pi i /N)$.

  The special case of $N=2$ arises in $S=1$ spin chains in which
  $\Integer_2\times\Integer_2$ is a subgroup of $SU(2)$ generated
  by $R^x = \exp(i\pi S^x)$ and $R^z=\exp(i\pi S^z)$. In these systems
  two phases can occur: a topological trivial and a non-trivial
  (Haldane) phase. The latter is characterized by a hidden symmetry
  breaking which becomes manifest after applying a non-local unitary
  transformation (NL-UT). \cite{raey}  The NL-UT can be written as
  \cite{0953-8984-4-36-019}
 \begin{align}\label{eq:NL-UToriginal}
  U_2=\prod_{i<j}\exp(\pi i S^x_iS^z_j)\ \ .
 \end{align}
  This transformation preserves the symmetry and maps local invariant
  Hamiltonians (such as that of the XYZ or the bi-linear bi-quadratic
  model) to local Hamiltonians. Most importantly, it maps the string
  order parameter $S^a_i \prod_{i\leq k<j}R^a_k S^a_j$ to a Landau
  order parameter $S^a_iS^a_j$ for $a = x$ or $z$, which explains that
  string order and hidden symmetry breaking are one-to-one related to
  each other.\footnote{Here and in what follows we deviate from the
    standard notation by letting the string extend to site $i$. This
    has the benefit of simplifying subsequent formulas.}

  Our attempt to generalize the previous considerations to
  $\Integer_N\times\Integer_N$ heavily used the string order selection
  rules introduced in Ref.~\onlinecite{PhysRevB.86.125441}. In order
  to explain the underlying ideas, let us consider a spin chain which
  is invariant under two commuting transformations $\FOP$ and
  $\SOP$. It is known that the action of these operators on the
  boundary modes can be factorized according to
\begin{align}
  \FOP\ =\ \FOP''\FOP'
  \quad\text{ and }\qquad
  \SOP\ =\ \SOP''\SOP'\ \ ,
\end{align}
  where the operators $\FOP''$ and $\SOP''$ act on the left boundary
  modes and $\FOP'$ and $\SOP'$ on the right boundary modes. The
  factorization leads to a phase ambiguity which implies that the
  boundary modes only need to transform projectively. In other words,
  we have that $\FOP'\SOP'=e^{i\phi}\SOP'\FOP'$ where the phase $\phi$
  determines the projective class of the representation of the right
  boundary mode. The phase $\phi$ thus also determines the topological
  phase of the system. Let us now consider a string order parameter
\begin{align}
 O^L_i\prod_{i\leq k<j}\FOP_k\,O^R_j\ \ .
\end{align}
  A non-vanishing expectation value of such a string order parameter
  in the limit $|i-j|\to\infty$ implies invariance under the
  transformation $\FOP$, but stated as such it contains no information
  on the topological phase. \cite{PhysRevLett.100.167202} However, the
  latter can be gained from a group theoretical selection
  rule. \cite{PhysRevB.86.125441} If the operators $O^L$ and
  $(O^R)^\dagger$ have the same quantum number with respect to $\SOP$,
\begin{align}
\SOP^{-1}O^L\SOP = e^{i\sigma}O^L \ \ \text{and} \ \ \SOP^{-1}O^R\SOP = e^{-i\sigma}O^R \ \ ,
\end{align}
  then the selection rule states that the string order parameter can
  only be nonzero if $\sigma = \phi$.

  For the case of systems with symmetry $\Integer_N\times\Integer_N$
  the role of $\FOP$ and $\SOP$ are played by $R$ and $\tilde{R}$,
  respectively. We define operators $X^a$ which are invariant under
  $R$ but which have a specific quantum number with respect to
  $\tilde{R}$:
\begin{align}\label{eq:propertiesX}
  \tilde{R}^{-1}X^a\tilde{R} = \omega^{-a}X^a\ \ ,  \ \ \ [R,X^a] = 0\ \ .
\end{align}
  Using these operators, we introduce the string order parameter
\begin{align}
 S(a,b) = X_i^a\prod_{i\leq k<j}R_k^b\,(X_j^a)^{-1}\ \ .
\end{align}
  We note that this operator becomes local for $b=0$. The selection
  rule for string order states that the expectation value
\begin{align}
 \Sigma(a,b) = \lim_{|i-j|\rightarrow\infty} \bigl\langle S(a,b)\bigr\rangle
\end{align}
  can only be nonzero if $a+tb=0$ modulo $N$, where $t$ is the
  projective class of the right boundary mode,
  \cite{PhysRevB.86.125441} see Eq.~\eqref{eq:define_t}.

  For the validity of our arguments below it is essential that all
  ground states of the system give rise to the {\em same}
  expectation value $\Sigma(a,b)$. This is due to the invariance of
  the string order parameter $S(a,b)$ under conjugation by $R$ and
  $\tilde{R}$. This invariance implies that all groundstates related
  by broken symmetries lead to the same result. On the other hand,
  groundstates in a well-defined gapped topological phase which are not
  related by such transformations only differ locally at the edges
  (see, e.g., Ref.\ \cite{PhysRevB.83.075102}), and the
  expectation value $\Sigma(a,b)$ is insensitive to such differences.
 
\section{\label{sec:NLTU}A non-local unitary transformation for \texorpdfstring{$\Integer_N\times\Integer_N$}{ZN x ZN}}

  In this section we aim to generalize the NL-UT given in
  Eq.~\eqref{eq:NL-UToriginal}, such that it is applicable to systems
  with $\Integer_N\times\Integer_N$ symmetry. Recall that this group
  is generated by the symmetries $R$ and $\tilde{R}$. Furthermore we
  consider two operators $O$ and $\tilde{O}$ which have the properties
  $R = \omega^O$ and $\tilde{R} = \omega^{\tilde{O}}$, with $\omega =
  \exp(2\pi i /N)$ as in the previous section. These operators will
  generalize $S^x$ and $S^z$. We then define a NL-UT as 
\begin{align}\label{eq:NL-UT}
U_N=\prod_{i<j}\omega^{{O}_i\tilde{O}_j}\ \ .
\end{align}
  All terms in the above product commute with each other. Commutators
  which are possibly nonzero are of the form
  $[\omega^{{O}_i\tilde{O}_j},\omega^{{O}_j\tilde{O}_k}]$. They can be
  rewritten as $[\tilde{R}_j^{{O}_{i}},R_j^{\tilde{O}_{k}}]$, from
  which it is clear that also these commutators are zero since $R$ and
  $\tilde{R}$ commute. Using similar arguments it is easily shown that
  both $R$ and $\tilde{R}$ commute with $U_N$. Thus $U_N$ preserves
  the $\Integer_N \times \Integer_N$ symmetry generated by these two
  transformations.

  Consider a $\Integer_N\times\Integer_N$ invariant local Hamiltonian
  $H_0$. We will now show that the transformed Hamiltonian $H_1 =
  U_N^{-1}H_0U_N$ is also local. More precisely, $n$-body terms (which
  act on $n$ consecutive sites) will be mapped to $n$-body terms. We
  will show this for $n=2$. Let $h_{i,i+1}$ be a term acting on sites
  $i$ and $i+1$. This term is transformed as follows:
\begin{align}\nonumber
U_N^{-1}h_{i,i+1}U_N &= P^\dagger
\omega^{-O_i\tilde{O}_{i+1}}h_{i,i+1}\omega^{O_i\tilde{O}_{i+1}}P\\ \label{eq:H2body}
& = \omega^{-O_i\tilde{O}_{i+1}}h_{i,i+1}\omega^{O_i\tilde{O}_{i+1}}\ \ ,
\end{align}
with
\begin{align}
P  = \prod_{j>i+1}(R_iR_{i+1})^{\tilde{O}_j} \prod_{j<i}(\tilde{R}_i\tilde{R}_{i+1})^{O_j}\ \ .
\end{align}
  The simplification in Eq.~\eqref{eq:H2body} is
  due to the $\Integer_N\times\Integer_N$ invariance of the
  Hamiltonian. The result is clearly a local 2-body term. The
  generalization to $n$-body Hamiltonians is straightforward.

  In the previous section it was explained that the string order
  parameter $S(a,b)$ is able to detect topological order via the
  selection rule. We will now discuss the transformation rule of
  $S(a,b)$. The operators $X^a$ appearing in $S(a,b)$ transform as
\begin{align}
 U_N^{-1}X^a_iU_N &= \prod_{j<i}\tilde{R}^{-O_j}_iX_i^a\prod_{j<i}\tilde{R}^{O_j}_i
 =\prod_{j<i}\omega^{-aO_j}X^a_i\ \ .
\end{align}
  In these equalities we have used Eq.~\eqref{eq:propertiesX}, in
  particular that $X^a$ and $R$ commute. With this transformation rule
  it follows that:
\begin{align} \nonumber
U_N^{-1}&S{(a,b)}U_N =  U_N^{-1}X^{a}_i\prod_{i\leq k <j}\omega^{bO_k}X_j^{-a}U_N & \\ \label{eq:string}
&= X^{a}_i\prod_{i\leq k <j}\omega^{(a+b)O_k}X^{-a}_j = S{(a,a+b)}\ \ .
\end{align}
  Applying $U_N$ sufficiently many times ($n$ times, such that
  $b+na=0$ modulo $N$), the result will eventually be
  $U_N^{-n}S{(a,b)}U_N^n = S{(a,0)}=X_i^aX_j^{-a}$. The operator
  $U_N^n$ thus relates the string order parameter which is capable of
  detecting topological phases to a Landau order parameter measuring
  symmetry breaking. Indeed, a nonzero $S{(a,b)}$ gives information
  on the topological phase through the selection rule, whereas a
  nonzero $S{(a,0)}$ gives information on the breaking of the
  symmetry generated by $\tilde{R}$.

  Just as before, we can define the string order parameter
  $\tilde{S}^{(a,b)} = \tilde{X}^a_i\prod_{i\leq
    k<j}\tilde{R}^b_k\tilde{X}^{-a}$ with operators $\tilde{X}^a$
  satisfying ${R}^{-1}\tilde{X}^a{R} = \omega^{a}\tilde{X}^a$ and
  $[\tilde{X}^a,\tilde{R}]=0$. Similar to Eq.~\eqref{eq:string} we
  have the transformation rule $U_N^{-1}\tilde{S}^{(a,b)}U_N =
  \tilde{S}^{(a,a+b)}$. In the topological phase labeled by $t$ both
  ${S}^{(a,b)}$ and $\tilde{S}^{(a,b)}$ can be nonzero (when their
  arguments satisfy $a+tb = 0$ mod $N$). If both string order
  parameters are zero, the unbroken symmetry transformations of the
  $U_N$-transformed system form a group of the form
  $\Integer_r\times\Integer_r\subset\Integer_N\times\Integer_N$.

\section{\label{sec:map}A mapping of phases}

  In this section we will discuss in detail what will happen to the
  symmetries after performing the NL-UT. That is, we start with a
  system with symmetry $\Integer_N\times\Integer_N$ whose ground
  states spontaneously break the symmetry to
  $\Integer_{r_0}\times\Integer_{r_0}$, with $q_0r_0 =
  N$. We furthermore assume that the system is in the topological
  phase $t_0$, defined by the projective class of the right boundary
  modes. In the previous section it was argued that the transformed
  system, obtained by applying the NL-UT defined in
  Eq.~\eqref{eq:NL-UT}, could show a different pattern of symmetry
  breaking and could reside in a different topological phase. It was
  also argued that the group of unbroken symmetries of the transformed
  system is of the form $\Integer_{r_1}\times\Integer_{r_1}$, with
  $q_1r_1 = N$. Let $t_1$ label the topological phase of the
  transformed system. We aim to find the explicit form of the relation
\begin{align}
 f_N: (r_0,t_0) \xrightarrow{\ \ U_N \ \ } (r_1,t_1)\ \ .
\end{align}

  As a warmup we will first assume that $N$ is prime. In this case,
  there is either no symmetry breaking or symmetry is fully
  broken. From the selection rule we conclude that only the string
  order parameter of the form $S{(at_0,-a)}$ can be nonzero. This
  string order parameter is mapped to $U_N^{-1}S{(at_0,-a)}U_N =
  S{(at_0,a(t_0-1))}$. As long as $t_0\neq1$ no symmetry is broken
  ($r_1 = N$). The topological phase can be deduced from the selection
  rule: $t_0 + t_1(t_0-1)=0$ mod $N$. In the exceptional case of
  $t_0=1$, the operator $S{(at_0,-a)}$ is mapped to a Landau order
  parameter measuring symmetry breaking ($r_1 = 1$ and trivially
  $t_1=0$). Conversely, if we start with a symmetry breaking phase,
  ($S{(a,0)} \neq 0$), the transformed system will have nonzero
  string order parameter $S{(a,a)}$ from which it follows that $t_1 =
  N-1$. Note that the trivial phase is always mapped to the trivial
  phase. In Table~\ref{fig:mapphase5} the action of the map $f_N$ is
  illustrated for $N=5$.

  The discussion is slightly more involved when $N$ is not prime
  because topological order can then be mixed with symmetry breaking
  order. The first step in the analysis is to use the selection rule
  to determine when $S(a,b)$ is possibly nonzero. Note that $b$ is
  always a multiple of $q_0$ ($b=n_1q_0$). The transformations $R^b$
  and $\tilde{R}^q$ restricted to the right boundaries do not commute
  but give rise to a phase factor. This complex phase depends on the
  topological phase $t_0$ via Eq.~\eqref{eq:define_t}:
\begin{align}
 \exp(2\pi i/r_0)^{t_0n_1} = \omega^{t_0b}\ \ .
\end{align}
  Moreover, from Eq.~\eqref{eq:propertiesX} it follows that
  transforming $X^a$ by $\tilde{R}^{q_0}$ gives rise to the phase
  $\omega^{-aq_0}$. The selection rule states that the string order
  parameter $S(a,b)$ is nonzero only if these two phases coincide,
  thus if $t_0b+q_0a = 0$ mod $N$. We conclude that nonzero string
  order parameters are of the form $S{(n_1t_0+n_2r_0,-n_1q_0)}$. An
  extra term $n_2r_0$ in the first argument is allowed since $X^{r_0}$
  commutes with $\tilde{R}^{q_0}$. Setting $n_1=0$ results in a
  nonzero Landau order parameter $S{(n_2r_0,0)}$, which is consistent
  with symmetry breaking at hand. This string order parameter is
  mapped to
\begin{align}\nonumber
 S(&n_1t_0+n_2r_0,-n_1q_0) \\ \label{eq:UnmapStoS}
 \xrightarrow{\ \ U_N \ \ } &S(n_1t_0+n_2r_0,n_1(t_0-q_0)+n_2r_0)\ \ .
\end{align}
  The transformed string order parameter is a Landau order parameter
  if its second argument vanishes (modulo $N$). This can only happen
  if $n_1(t_0-q_0) = 0$ mod $r_0$. The smallest $n_1$ which fulfills
  this equation is given by $n_1 = r_0/$gcd$(t_0-q_0,r_0)$. The
  corresponding symmetry breaking operator $X^a$ is determined by $a =
  n_1t_0+n_2r_0 = n_1q_0 = N/$gcd$(t_0-q_0,r_0)$. We conclude that the
  symmetry of the transformed system is determined by
\begin{equation}
 r_1 = \frac{N}{\text{gcd}(t_0-q_0,r_0)}\ \ .
\end{equation}
  Note that the second argument of the transformed string operator is
  a multiple of $q_1 = \text{gcd}(t_0-q_0,r_0)$. Thus the selection
  rule can be used to determine the topological phase $t_1$ of the
  transformed system. From this rule we have that
\begin{equation}
 (n_1t_0+n_2r_0)q_1 = -t_1\bigl[n_1(t_0-q_0)+n_2r_0\bigr]\ \ .
\end{equation}
  The solution for $t_1$ should be independent of $n_1$ and
  $n_2$. Thus factoring these constants out we obtain two equations
\begin{align}
 0 &= t_1(t_0-q_0)+t_0q_1\ \ \text{mod}\ \ N \ \ ,\\
  0 &= r_0(q_1+t_1)\ \ \text{mod}\ \ N \ \ .
\end{align}
  From the second equation we deduce that $t_1$ is equal to $-q_1$
  modulo $q_0$ thus $t_1 = nq_0-q_1$. Substituting in the first and
  simplifying results in $n(t_0-q_0) = -q_1$ mod $r_0$. This equation
  has a solution for $n$ due to the definition of $q_1
  =\gcd(t_0-q_0,r_0)$. In Table~\ref{fig:mapphase5} the map $f_N$ is
  worked out for $N=6$.

\begin{table}
\center
 \begin{tabular}[t]{ccc}\hline \hline
  &$\Integer_5\times\Integer_5$&\\
  $t_0$&&$t_1$\\\hline \hline
  $0$&$\rightarrow$&$0$\\
  $1$&$\rightarrow$&SSB\\
  $2$&$\rightarrow$&$3$\\
  $3$&$\rightarrow$&$1$\\
  $4$&$\rightarrow$&$2$\\
  SSB&$\rightarrow$&$4$\\\hline \hline
 \end{tabular}
 \qquad\qquad
  \begin{tabular}[t]{ccc}\hline \hline
  &$\Integer_6\times\Integer_6$&\\
  $(r_0,t_0)$&&$(r_1,t_1)$\\\hline \hline
  $(6,0)$&$\rightarrow$&$(6,0)$\\
  $(6,1)$&$\rightarrow$&$(1,0)$\,SSB\\
  $(6,2)$&$\rightarrow$&$(6,4)$\\
  $(6,3)$&$\rightarrow$&$(3,0)$\\
  $(6,4)$&$\rightarrow$&$(2,0)$\\
  $(6,5)$&$\rightarrow$&$(3,2)$\\ \hline
  $(3,0)$&$\rightarrow$&$(6,3)$\\
  $(3,1)$&$\rightarrow$&$(6,1)$\\
  $(3,2)$&$\rightarrow$&$(2,1)$\\ \hline
  $(2,0)$&$\rightarrow$&$(6,2)$\\
  $(2,1)$&$\rightarrow$&$(3,1)$\\ \hline
  $(1,0)$\,SSB&$\rightarrow$&$(6,5)$
  \\\hline  \hline
 \end{tabular}
 \caption{Mapping of symmetry breaking and topological phases of a $\Integer_5\times\Integer_5$ (left) and a $\Integer_6\times\Integer_6$ (right) invariant system. With SSB we refer to the phase characterized by full spontaneous symmetry breaking.}
 \label{fig:mapphase5}
\end{table}

  Let us finally comment on a subtle technical issue. Our derivation
  of the map $f_N$ hings on the presence of (string) order via the
  selection rule. However, the selection rule only gives a necessary
  but not a sufficient condition on the non-vanishing of the (string)
  order. It can be accidentally absent at specific points in the phase
  diagram where one would have expected it to occur from the selection
  rule. To resolve this problem, we recall that the map $f_N$ is not a
  statement about specific points in a phase but rather about phases
  as a whole. Assuming the existence of some point in the phase
  diagram where the groundstate leads to a non-vanishing string order
  is enough to show that the map $f_N$ is valid for the whole phase.

  Note that from Table~\ref{fig:mapphase5} it can be seen that $f_N$
  is bijective for $N=5$ and $N=6$. This is as expected since the
  spectrum and thus phase transitions should be invariant under the
  NL-UT. Also it has been checked numerically for values of $N$ up to
  100 that indeed $f_N$ is bijective.

  In Ref.~\onlinecite{Else:2013arXiv1304.0783E} a different NL-UT
  transformation, mapping topological phases to symmetry breaking
  phases, is discussed. The main difference is that we discuss one
  single NL-UT whereas in Ref. \onlinecite{Else:2013arXiv1304.0783E} the NL-UT
  $\mathcal{D}_t$ depends on the topological phase $t$ at hand. The
  map $\mathcal{D}_t$ always maps a pure topological phase to the
  phase where the full symmetry is spontaneously broken (SSB-phase),
  whereas $U_N$ maps phases characterized by a mixture of symmetry
  breaking and topology to each other. Under certain conditions one
  can map such phases to the SSB-phase applying the transformation
  $U_N$ sufficiently many times. Indeed, from Eq.~\eqref{eq:UnmapStoS}
  we have that
\begin{align} \nonumber
  S(&n_1t_0+n_2r_0,-n_1q_0)  \\
  \xrightarrow{\ \ U^u_N \ \ }&S(n_1t_0+n_2r_0,n_1(ut_0-q_0)+n_2ur_0)\ \ ,
\end{align}
  where $u$ is the number of times $U_N$ is applied. An arbitrary
  phase is mapped to the SSB-phase if its string order parameters are
  mapped to Landau order parameters, that is if
  $n_1(ut_0-q_0)+n_2ur_0=0$ mod $N$ for all values of $n_1$ and
  $n_2$. From this we have the conditions that $ut_0=q_0$ mod $N$ and
  $u=0$ mod $q_0$. The first condition can only be satisfied if
  gcd($t_0,r_0) = 1$. This is equivalent to stating that the
  projective representations of $\Integer_{r_0}\times\Integer_{r_0}$
  in class $t_0$ are \textit{maximally non-commutative},
  \cite{Else:2013arXiv1304.0783E} a requirement for defining the map
  $\mathcal{D}_{t_0}$. The  second condition can be verified by
  considering orbits of $f_N$ containing the SSB-phase. For example,
  taking $N=6$ this orbit is
\begin{align}\nonumber
 \text{SSB} \rightarrow (6,5) \rightarrow (3,2) \rightarrow (2,1) \rightarrow (3,1) \rightarrow (6,1) \rightarrow \text{SSB}\ \ .
\end{align}

\section{\label{sec:continuoussymmetrygroups}Relevance to systems with continuous symmetry groups}

  Although the previous section gives a better understanding of the
  connection between symmetry breaking order and topological order, as
  it stands it is only applicable to systems with the specific
  symmetry $\Integer_N\times\Integer_N$. In this section, however, we
  will argue that almost all connected and compact simple Lie groups
  $G_\Gamma$ (with only one exception) contain a subgroup of the form
  $\Integer_N\times\Integer_N$ which is sensitive to the projective
  classes of $G_\Gamma$. More precisely, we will be discussing
  subgroups $F_\Gamma \subset G_\Gamma$ such that the homomorphism
\begin{align}
\tau:\ H^2\bigl(G_\Gamma,U(1)\bigr)\rightarrow H^2\bigl(F_\Gamma,U(1)\bigr)
\end{align}
  is bijective and use a case by case argument to show that $F_\Gamma$
  can be chosen to be of the form
  $\Integer_N\times\Integer_N$. Topological order can thus also be
  understood as ``hidden'' symmetry breaking if it is protected by a
  continuous symmetry. Note that this is not in contradiction with the
  Mermin-Wagner Theorem \cite{PhysRevLett.17.1133} (stating that
  spontaneous symmetry breaking does not occur in one-dimensional
  quantum systems with continuous symmetry) since the NL-UT only
  preserves the discrete subgroup $\Integer_N\times\Integer_N$ but not
  the full continuous symmetry.

  As was explained in great detail in
  Ref.~\onlinecite{Duivenvoorden:2012arXiv1206.2462D}, a connected compact
  simple Lie group can be written as a a quotient $G_\Gamma =
  G/\Gamma$ (which motivates the notation) where $G$ is the universal
  cover of $G_\Gamma$ and $\Gamma$ is a subgroup of the center
  $\cZ(G)$ of $G$. All projective representations $\rho$ of $G_\Gamma$
  originate from linear representations of $G$. The projective class
  of a representation of $G_\Gamma$ can be deduced from the action of
  $\Gamma$ on the corresponding linear representation of $G$ (denoted
  by $\rho: \Gamma \rightarrow U(1)$). With only one exception
  (occuring for $G=Spin(4n)$), $\Gamma$ is of the form
  $\Integer_N$. In these cases we choose a generator
  $\gamma\in\Gamma$ and define the projective class of a
  representation by $\rho(\gamma) =\omega^t$ with $\omega = \exp(2\pi
  i/N)$. The main strategy in defining $F_\Gamma$ is to start by
  choosing $R$, $\tilde{R}\in G$ such that $R^N \in \Gamma$ and
  $\tilde{R}^N\in\Gamma$ (moreover, $N$ should be the smallest nonzero
  integer for which this holds). Furthermore,
  $R\tilde{R}R^{-1}\tilde{R}^{-1}$ should generate $\Gamma$. Thus
  $R\tilde{R}R^{-1}\tilde{R}^{-1} = \gamma^m$ with gcd$(N,m)=1$. Let
  $F$ be generated by $R$ and $\tilde{R}$. Clearly $\Gamma\subset
  F$. Furthermore $F/\Gamma = F_\Gamma = \Integer_N\times\Integer_N$
  is a finite Abelian group.

  It is now not to hard to see that with this choice of $F_\Gamma$ the
  map $\tau$ is bijective. Let $\rho$ be a representation of $G$ in
  the projective class $t$ ($\rho(\gamma)=\omega^t$ for $\gamma\in
  \cZ(G)$). Restrict this representation to $F$. The projective class of
  this restricted representation is determined by the phase obtained
  upon commuting $R$ and $\tilde{R}$ or, in other words, by
  $\rho(\gamma^m) = \omega^{tm}$. As a consequence we find $\tau:
  t\rightarrow tm$ which is bijective under the assumption that
  gcd$(N,m)=1$.

  It remains to show that symmetry generators $R$ and $\tilde{R}$
  with the desired properties indeed exist. In what follows, we will
  provide an explicit realization for $G$ being equal to either of the
  groups $SU(N)$, $Sp(N)$ or $Spin(N)$. A summary of the possible
  quotient groups to be considered can be found in Ref.~\onlinecite{Duivenvoorden:2012arXiv1206.2462D}.

\paragraph{Case $G_\Gamma = SU(N)/\Integer_N$:} The subgroup
$\Integer_N$ is generated by $\omega\mathbb{I}$, where $\omega =
\exp(2\pi i /N)$. We will give matrix representations of $R$ and
$\tilde{R}$ acting on $\mathbb{C}^N$. Let $v_i$ denote the standard
orthonormal basis of this space. Choose $R$ to be proportional to the
linear map $v_i \rightarrow \omega^iv_{i}$. The proportionality
constant $c$ should be chosen such that $R$ has determinant 1, thus
$c^N=(-1)^{N+1}$. Similarly, let $\tilde{R}$ be proportional to the
linear map $v_i \rightarrow v_{i+1}$ (and $v_N \rightarrow v_{1}$)
with the same proportionality constant $c$. These matrices obey $R^N =
\tilde{R}^N = (-1)^{N+1}\mathbb{I}$ and $R\tilde{R} =
\omega\tilde{R}R$.

\paragraph{Case $G_\Gamma =SU(N)/\Integer_q$:} The subgroup
$\Integer_q$ is generated by $\omega\mathbb{I}$, where $\omega =
\exp(2\pi i /q)$. Choose the generators of $F$ to be the block
diagonal matrices $R=\diag(R_q,\ldots,R_q)$ and
$\tilde{R}=\diag(\tilde{R}_q,\ldots,\tilde{R}_q)$ where $R_q$ and
$\tilde{R}_q$ are defined just as for the case $SU(q)/\Integer_q$
discussed before. From the previous paragraph we then directly
conclude that $R^q = \tilde{R}^q = (-1)^{q+1}\mathbb{I}$ and
$R\tilde{R} = \omega\tilde{R}R$.

\paragraph{Case $G_\Gamma =Sp(2N)/\Integer_2$:} The group $Sp(2N)$
consists of complex unitary $2N\times2N$ matrices $M$ which preserve a
symplectic form $Q$ (i.e.\ $M^TQM = Q$ where $Q$ is non-singular skew
symmetric matrix). Choose a basis such that $Q  =
\bigl(\begin{smallmatrix}0&-\mathbb{I}\\
  \mathbb{I}&0 \end{smallmatrix} \bigr)$. The group
$\Integer_2$ is $\{\mathbb{I}, -\mathbb{I}\}$. Let the generators of
$F$ be $R  = \bigl(\begin{smallmatrix} i\mathbb{I}&0\\ 0&-i \mathbb{I}
\end{smallmatrix} \bigr)$, and $\tilde{R}  = Q$. Clearly $R^2 =
\tilde{R}^2 = -\mathbb{I}$ and $R\tilde{R}= -\tilde{R}R$.

\paragraph{Case $G_\Gamma =Spin(N)/\Integer_2 = SO(N)$, $N\geq3$:} The
group $Spin(N)$ is most easily understood by first considering its Lie
algebra $so(N)$. Let $e_i$ be an orthonormal basis of
$\mathbb{R}^N$. The Clifford algebra $\Cl(N)$ is generated by this
basis together with the relations $\{e_i,e_j\}=2\delta_{ij}$. The Lie
algebra $so(N)\subset\Cl(N)$ is generated by the
operators $S_{ij} = \frac{i}{2}e_ie_j$ (with $i\neq j$). These
operators generate rotations in the $(i,j)$-plane: $T_{ij}(\theta) =
e^{i\theta
  S_{ij}}=\cos(\frac{\theta}{2})\mathbb{I}-\sin(\frac{\theta}{2})e_ie_j$.
  The group $\Integer_2$ is $\{\mathbb{I}, -\mathbb{I}\}$. The
  generators of $F$ can be chosen to be $R = e_1e_2$ and
  $\tilde{R}=e_2e_3$. These elements square to $-\mathbb{I}$ and
  anti-commute. From the perspective of $SO(N)$, they correspond to
  $\pi$-rotations in two orthogonal planes with a one-dimensional
  intersection.

\paragraph{Case $G_\Gamma =Spin(N)/\Integer_4=PSO(N)$, $N=4n+2$:} In
this case the center of $Spin(N)$ is isomorphic to $\Integer_4$ and is
generated by the element $\gamma = \prod_ie_i$ (indeed $\gamma^2 =
-\mathbb{I}$). The generators of $F$ can be chosen to be $R =
2^{-1/2}(1+e_{N-1}e_N)\prod_{i=1}^{n} e_{4i-2}e_{4i}$ and $\tilde{R} =
2^{-n}e_{N-2}e_N \prod_{i=1}^{2n} (1+e_{2i-1}e_{2i})$. One can check
that $R^4=\tilde{R}^4=-\mathbb{I}$ and that $R\tilde{R} = \gamma
\tilde{R}R$.

\paragraph{Case $G_\Gamma =Spin(N)/(\Integer_2\times\Integer_2) =
  PSO(N)$, $N=4n$:} This is the only exception to the above recipe
since the center of $Spin(N)$ is no longer of the form $\Integer_q$ if
$N$ is a multiple of four.  However we can still give a flavor of how
this exception should be treated. Again, let $\gamma =
\prod_ie_i$. The center is generated by $\gamma$ and $-\mathbb{I}$
(indeed $\gamma^2 = \mathbb{I}$). The projective class of a
representation $\rho$ is determined by both $\rho(\gamma) =
\pm\mathbb{I}$ and by $\rho(-\mathbb{I})=\pm\mathbb{I}$. In this case
we shall thus need two string order parameters to determine both
pre-factors via the selection rule. One could define two different
NL-UTs mapping these string order parameters to Landau order
parameters. In order to define both the string order parameters as
well as the NL-UTs, one would need to go through the above procedure
twice. That is, define $R_i$, $\tilde{R}_i\in Spin(N)$ for $i=1,2$
such that $R_1\tilde{R}_1 = -\tilde{R}_1R_1$ and $R_2\tilde{R}_2 =
\gamma \tilde{R}_2R_2$. Let $R_1 = R_2  = \prod_{i=1}^{n}
e_{4i-2}e_{4i}$, $\tilde{R}_1 = e_1e_2$ and
$\tilde{R}_2 =  2^{-n}\prod_{i=1}^{2n} (1+e_{2i-1}e_{2i})$, which satisfy
the desired conditions (together with ${R}_1^2 = (-1)^{n}\mathbb{I}$,
$\tilde{R}_1^2 = -\mathbb{I}$ and $\tilde{R}_2^2 = \gamma$).

\paragraph{Case $G_\Gamma =Spin(N)/\Integer_2 = SS(N)$, $N=4n$:} As
discussed before, the group $Spin(N)$ has center
$\{\mathbb{I},-\mathbb{I},\gamma,-\gamma\}=
\Integer_2\times\Integer_2$. After dividing out $\{\mathbb{I},\gamma\}
$ or $\{\mathbb{I},-\gamma\}$ (which are both isomorphic to
$\Integer_2$) one obtains the group $SS(N)$ also known as the
semi-spinor group (see, e.g.,
Ref.~\onlinecite{Duivenvoorden:2012arXiv1206.2462D}). In the former
case one could define $R =
R_2$ and $\tilde{R} =\tilde{R}_2$, where  $R_2$ and $\tilde{R}_2$ have
been defined in the previous paragraph. Using the same reasoning as
before, we directly obtain $R\tilde{R} = \gamma \tilde{R}R$ and
$\tilde{R}^2 = \gamma\mathbb{I}$. However, $R^2=-\mathbb{I}\notin \Gamma$. Thus
although $F_\Gamma = \Integer_4\times\Integer_2$ constructed in this
way does lead to an bijective $\tau$, it is not of the form
$\Integer_r\times\Integer_r$. In principle the results discussed in
Sect.~\ref{sec:NLTU} and \ref{sec:map} can not be directly
applied. However, Eq.~\eqref{eq:NL-UT} can still be used to define a
NL-UT and an analysis similar to Sect.~\ref{sec:map} can be performed
to find out what type of symmetry breaking phase the topological
non-trivial phase is mapped to by such a NL-UT. In the case where $\Gamma
= \Integer_2 =\{\mathbb{I},-\gamma\}$ similar problems occur. One
could still define $R = R_2$ and $\tilde{R} =e_1e_2\tilde{R}_2$ such
that $R\tilde{R} = -\gamma \tilde{R}R$ and $\tilde{R}^2 =
-\gamma\mathbb{I}$. But also in this case $R^2=-\mathbb{I}\notin
\Gamma$ such that $F_\Gamma = \Integer_4\times\Integer_2$.

\section{Conclusions and outlook}

  We presented a non-local unitary transformation which maps
  topological phases of $\Integer_N\times\Integer_N$ spin chains to
  symmetry breaking ones. Since the map transfers non-local string
  order to the more familiar local Landau order it provides a useful
  alternative characterization of topological phases. Our result may
  be regarded as a two-fold generalization of the ``hidden'' symmetry
  breaking that is familiar from the AKLT model. \cite{raey} First of
  all, our method is able to deal with the existence of several distinct
  non-trivial topological phases, not just a single one. Moreover, in
  view of the existence of non-trivial subgroups
  $\Integer_r\times\Integer_r\subset\Integer_N\times\Integer_N$, we
  are also capable of characterizing phases which exhibit a mixture of
  topological protection and spontaneous symmetry breaking.

  As pointed out in Sect.~\ref{sec:continuoussymmetrygroups}, our
  previous considerations lead to
  a full characterization of topological phases in spin systems with
  continuous symmetry groups. This observation relies on the existence
  of discrete subgroups of type $\Integer_N\times\Integer_N$ in all
  classical groups of unitary, orthogonal or symplectic type. Besides
  constructing these subgroups explicitly, we also showed that the
  projective representations of the continuous groups and their
  subgroups are (with one exception) in one-to-one correspondence if
  $N$ is chosen properly. For the stability of edge modes it is thus
  not important to preserve the full continuous symmetry group. Rather
  it is sufficient to preserve the corresponding discrete
  subgroup. This phenomenon has been known for some time in the case
  of $SO(3)$
  \cite{Pollmann:PhysRevB.81.064439,Pollmann:2012PhRvB..85g5125P} but
  the picture that emerges from our paper is somewhat more complete.

  As a by-product, our analysis provides a complementary perspective
  on the hierarchy of topological phases that was pointed out in
  Ref.~\onlinecite{Duivenvoorden:2012arXiv1206.2462D}. As was shown in
  Ref.~\onlinecite{Duivenvoorden:2012arXiv1206.2462D}, there is an injection
  of topological phases when viewing the same system from the
  perspective of either $G_\Gamma$ or $G_{\Gamma'}$, with
  $\Gamma'\subset\Gamma\subset\cZ(G)$ being two central subgroups of $G$. In
  Sect.~\ref{sec:continuoussymmetrygroups} we have shown that
  $G_\Gamma$ has a subgroup $F_\Gamma$ that can be used to
  characterize the topological phases, and a similar statement holds
  for $\Gamma'$. From the construction of $F_\Gamma$ it is clear that
  $F_{\Gamma'}\subset F_\Gamma$. The original hierarchy of topological
  phases is thus also reflected on the level of ``hidden'' symmetry
  breaking.

  We would like to stress that our results from
  Sect.~\ref{sec:continuoussymmetrygroups} also provide a precise
  route to embed spin systems with discrete spin degrees of freedom into spin
  systems with continuous symmetry. While \textit{a priori} the latter appear
  to be more constrained, this reformulation may nevertheless be
  useful with regard to constructing effective low energy topological
  field theories in terms of non-linear $\sigma$-models. In this
  sense, it may be used to make some of the ideas discussed in
  Ref.~\onlinecite{Chen:2011arXiv1106.4772C} more precise. It remains to be
  clarified, however, whether the embeddings just mentioned only
  capture the features of topological protection with regard to
  continuous symmetry groups in one-dimensional systems or whether the
  correspondence also lifts to higher dimensions.

\begin{center}
  \em $\ast\ast\ast$\quad Note\quad $\ast\ast\ast$
\end{center}
  \vspace{-.8em}
Recently Ref.~\onlinecite{Else:2013arXiv1304.0783E} appeared which has considerable
  overlap with our own results. Let us briefly summarize the main
  differences. Our setup is, in a sense, more limited. Instead of
  considering arbitrary Abelian groups, we restrict our attention to
  groups of type $\Integer_N\times\Integer_N$. Besides, the authors of
  Ref.~\onlinecite{Else:2013arXiv1304.0783E} construct a different non-local
  unitary transformation for each individual purely topological phase
  which can be used to map it to a phase with {\em full} spontaneous
  symmetry breaking. In contrast, we keep the non-local unitary
  transformation fixed and investigate the fate of various phases
  under this map. This allows us to investigate phases which
  involve a mixture of topological protection and symmetry
  breaking. At the end of Sect.~\ref{sec:map} we determine precise
  conditions under which a given phase can possibly be mapped to a
  fully symmetry broken one. Finally, our treatment of continuous
  groups exhausts {\em all} classical cases and does not just cover
  $PSU(N)$ and $SO(2N+1)$ as in Ref.~\onlinecite{Else:2013arXiv1304.0783E}.

\begin{acknowledgments}
  We gratefully acknowledge useful and interesting discussions with
  R.\ Bondesan, N.\ Schuch, A.\ Turner and J.\ Zaanen. The work of K.\
  Duivenvoorden is funded by the German Research Foundation (DFG)
  through the SFB$|$TR\,12 ``Symmetries and Universality in Mesoscopic
  Systems'' and the ``Bonn-Cologne Graduate School of Physics and
  Astronomy'' (BCGS). The work of T.\ Quella is funded by the DFG
  through M.\ Zirnbauer's Leibniz Prize, DFG Grant No.\ ZI 513/2-1.
\end{acknowledgments}


%

\end{document}